\documentclass[10pt,twocolumn,showpacs,preprintnumbers,amsmath,amssymb,aps,prl,longbibliography,superscriptaddress]{revtex4-2}

\usepackage[utf8]{inputenc}
\usepackage{graphicx}
\usepackage{dcolumn}
\usepackage{bm}
\usepackage{hyperref}
\usepackage{float}
\usepackage{cleveref}
\crefname{appendix}{App.}{Apps.}
\crefname{equation}{Eq.}{Eqs.}
\crefname{figure}{Fig.}{Figs.}
\crefname{table}{Tab.}{Tabs.}
\crefname{section}{Sec.}{Secs.}

\usepackage{xcolor}  
\definecolor{lightgray}{gray}{0.6}
\definecolor{medgray}{gray}{0.4}
\definecolor{mypink}{rgb}{0.858, 0.188, 0.478} 

\newif\ifptitle
\newif\ifpnumber
\newcounter{para}

\ptitletrue  
\pnumbertrue  

\begin{document}

\title{Spin Chain with Continuous Exponential Symmetry: Topological Phase and Beyond}

\author{Biao Lian}
\altaffiliation{biao@princeton.edu}
\affiliation{Department of Physics, Princeton University, Princeton, New Jersey 08544, USA}
\author{Zhen Bi}
\altaffiliation{zjb5184@psu.edu}
\affiliation{Department of Physics, The Pennsylvania State University, University Park, Pennsylvania 16802, USA} \affiliation{Center for Theory of Emergent Quantum Matter$,$ Institute for Computational and Data Sciences$,$ The Pennsylvania State University$,$ University Park$,$ Pennsylvania$,$ 16802$,$ USA}
\date{\today}

\begin{abstract}
We study symmetric gapped ground states of one-dimensional spin $S$ chains with a continuous exponential symmetry $U(1)\rtimes\mathbb Z_2$ with integer base $q$, where the U(1) charge has a modulated unit of charge $q^{L-j}$ decaying exponentially with site $j$. The simplest model is a quantum breakdown model that can be viewed as a $q$-deformation of the XXZ Haldane spin chain. We show that the topology of their symmetric ground states crucially depends on $S$ and $q$. For integer $S$, odd $q$ allows a nontrivial $\mathbb Z_2$ Haldane-type symmetry-protected topological (SPT) phase with protected edge modes, while even $q$ allows no nontrivial SPT phase. For half-odd integer $S$, odd $q$ prohibits any unique symmetric gapped ground state due to a Lieb-Schultz-Mattis obstruction, whereas even $q$ enforces a novel symmetric gapped phase which we name as the ``half-SPT" phase. Unconventionally, the half-SPT phase only has a symmetry protected edge mode on one edge, and has half of the string orders of an SPT phase. Our results uncover a new type of symmetry-protected gapped phases beyond existing classification schemes.
\end{abstract}

\maketitle

Symmetries in quantum many-body systems can protect topologically nontrivial ground states, known as the symmetry protected topological (SPT) phases\cite{Chen2011Classification,Schuch2011Classifying,Chen2013GroupCohomology,Gu2009SPT,Chen2012BosonicSPT,Senthil2015SPT}. Seminal examples include one-dimensional (1D) Haldane spin chain with spin rotational symmetry \cite{haldane1983continuum,haldane1983nonlinear,AKLT1987,Pollmann2012SymmetryProtection}, and topological insulators with time-reversal symmetry \cite{Hasan2010,Qi2011}. A universal feature of the SPT phases is the existence of symmetry-protected topological edge states. In 1D, an SPT phase with symmetry group $G$ is defined to have (i) a unique symmetric gapped ground state in periodic boundary condition (PBC), and (ii) a zero-energy edge mode per edge carrying a projective irreducible representation (irrep) of $G$ in open boundary condition (OBC)\cite{Chen2011Classification,Schuch2011Classifying}. In the example of Haldane phase of the anti-ferromagnetic (AFM) spin-1 chain with SO(3) or $\text{U}(1)\rtimes\mathbb{Z}_2$ symmetry \cite{haldane1983continuum,haldane1983nonlinear}, each edge in OBC has a zero-energy spin-$\frac{1}{2}$ edge mode, which can be solved exactly in the Affleck-Kennedy-Lieb-Tasaki (AKLT) limit \cite{AKLT1987}.

In recent years, modulated symmetries \cite{Sala_2022_dynamics} in lattice systems, which feature a spatially varying unit of symmetry charge and thus do not commute with lattice translations, have attracted growing interests. Examples include the dipole or multipole symmetry \cite{bergholtz2008,nakamura2012,Lake2022dipolar,schulz2019stark,Khemani2020localization,sala2020,Gorantla2022dipole,Ye2020fractonic,Ye2021fractonic,lake2022multipole,Ye2025fractonic}, harmonically modulated symmetry and exponential symmetry \cite{lian2023quantum,chen2024quantum,liu20232d,hu2025bosonic,hu2026spin,Kaplan_2016,hu_watanabe_2023,watanabe2023ground,delfino2023,Sala2024exotic,wang2025exponentiallyslowthermalization1d,pace2024}. The composition of these three types of symmetries gives the generic modulated symmetries in translationally invariant systems \cite{chenbt2026,yoshitome2026}. It has been shown that discrete modulated symmetries, including those realized in clock models, can protect
SPT phases\cite{Saito_2025,Han_2023fas,Lam_classification,Kim_2025,Yao2025iia,ning2026matrixproductstatesmodulated,bulmash2025,anakru2026matrixproductstatesmodulated,Jones2025ChargePumps,Lee2026DipolarCluster}. However, SPT phases protected by continuous modulated
symmetries and their realization in quantum spin models
remain underexplored. Among such symmetries, the exponential U(1) symmetry is particularly interesting, which has a unit of charge $q^{L-j}$ exponentially decaying in site $j$ of a 1D chain of length $L$. Spin chains with this symmetry, known as quantum breakdown models, can exhibit a highly unconventional spontaneous symmetry breaking (SSB) phase with $\mathcal{O}(q^L)$ number of ground states (one per charge sector), which resembles a spin glass \cite{hu2025bosonic,hu2026spin}, albeit in a translationally invariant system. 

In this letter, we study spin $S$ chain with an exponential $\text{U}(1)\rtimes\mathbb{Z}_2$ symmetry with integer base $q$, which gives a quantum breakdown model that can be viewed as a $q$-deformation of the XXZ Haldane spin chain (see \cref{eq-H}). We find their symmetric gapped ground states crucially depend on spin $S$ and integer $q$ (see \cref{table:classification}). For integer $S$, a nontrivial $\mathbb{Z}_2$ SPT phase similar to the Haldane phase can arise for odd $q$, and no nontrivial SPT phase exists for even $q$. For half-odd integer $S$, odd $q$ leads to a Lieb-Schultz-Mattis (LSM) obstruction to a unique symmetric gapped ground state, while even $q$ enforces a novel symmetric ground state which we term as a ``half-SPT". In the half-SPT phase, the ground state is unique in PBC, and the symmetry protects a zero-energy edge mode only at the right edge in OBC. Also, it  possesses half of the long-range string orders of the usual SPT, and has the same entanglement-spectrum degeneracy as the usual SPT. This reveals a new type of symmetric gapped phases beyond the existing classification scheme.


\emph{Symmetry and Model}---We consider a 1D spin chain of length $L$ with a spin $S$ per site, where the spin operators on site $j$ are $\hat S^\mu_j$ ($\mu=x,y,z$). In OBC, we assume the Hamiltonian has a continuous \emph{exponential symmetry} group $G=\text{U}(1)\rtimes\mathbb{Z}_2$, with group elements given by $G=\{g(\nu,\varphi)=\hat X^\nu e^{i\varphi \hat Q}|\nu\in\{0,1\},\varphi\in 2\pi\mathbb{R}/\mathbb{Z}\}$, where the exponential $\text{U}(1)$ charge $\hat Q$ and $\mathbb{Z}_2$ generator $\hat X$ are defined as
\begin{equation}\label{eq-exp-charge}
\begin{split}
\hat Q=\sum_{j=1}^L q^{L-j}\hat S_j^z\ ,\qquad \hat X=\prod_{j=1}^{L}\exp\left(i\pi \hat S_x^j\right)\ .
\end{split}
\end{equation}
In this study, we restrict $q\in\mathbb{Z}_+$ (denoting positive integers), so that we can identify $\varphi\simeq\varphi+2\pi$. When $q=1$, $G$ reduces to the conventional symmetry group of the XXZ model. In PBC, $g(\nu,\varphi)$ must rotate the spins on the $0$-th site and the $L$-th site by the same angle, which requires $q^L\varphi=\varphi\ (\text{mod }2\pi)$. Thus, for $q\ge 2$, the symmetry group $G$ is reduced to a discrete group $G_{L}=\mathbb{Z}_{q^L-1}\rtimes\mathbb{Z}_2\subset G$ in PBC. The boundary condition dependence of the exponential symmetry is a crucial difference compared to conventional symmetries.

For $q\ge 2$, as summarized in the supplementary material (SM) \cite{suppl}, the OBC exponential symmetry group $G=\text{U}(1)\rtimes\mathbb{Z}_2$ has two linear (i.e. non-projective) 1D singlet irreps $\mathcal{R}^\text{1D}_\pm(g(\nu,\varphi))=(\pm1)^\nu$, and infinitely many 2D irreps $\mathcal{R}^\text{2D}_{m_z}(g(\nu,\varphi))= (\sigma_x)^\nu \exp\left(i m_z\varphi\sigma_z\right)$ labeled by a positive \emph{pseudospin} $m_z\in\mathbb{Z}_+/2$, where $\sigma_\mu$ ($\mu=x,y,z$) are Pauli matrices. The 2D irrep $\mathcal{R}^\text{2D}_{m_z}$ is linear if $m_z\in\mathbb{Z}_+$, and projective if $m_z\in \mathbb{Z}_+-\frac{1}{2}$. With charge $\hat Q$ in \cref{eq-exp-charge}, the pseudospin $m_z$ differs from the physical $z$-direction spin by a site-dependent weight $q^{L-j}$. The PBC group $G_L$ has similar irreps, except that its 2D irreps are restricted to $0<m_z<\frac{q^L-1}{2}$, and it has two additional 1D irreps $\widetilde{\mathcal{R}}^\text{1D}_\pm(g(\nu,\varphi))=(\pm1)^\nu e^{i(q^L-1)\varphi/2}$ reduced from the $m_z=\frac{q^L-1}{2}$ 2D irrep of $G$ (see SM \cite{suppl}).


While our study of symmetric gapped ground states do not rely on specific Hamiltonians, we will mainly verify our conclusions numerically in the spin quantum breakdown Hamiltonian with symmetry group $G$ below:
\begin{equation}\label{eq-H}
\begin{split}
&\hat H=\sum_j\left[ \hat {S}_{ j}^{+}(\hat {S}_{ j+1}^{-})^q + \hat {S}_{ j}^{-}(\hat {S}_{ j+1}^+)^q +V \hat {S}_{ j}^{z}\hat {S}_{ j+1}^{z}\right]\ ,
\end{split}
\end{equation}
where $\hat{S}_j^\pm =\hat{S}_j^x\pm i\hat{S}_j^y$, the spin $S\ge\frac{q}{2}$ is required for the first two terms to be nonzero, and $V$ is a real parameter. This model can be viewed as an inversion-breaking $q$-deformation of the XXZ Haldane spin chain.

\begin{figure}[t]
\centering
\includegraphics[width=3.5in]{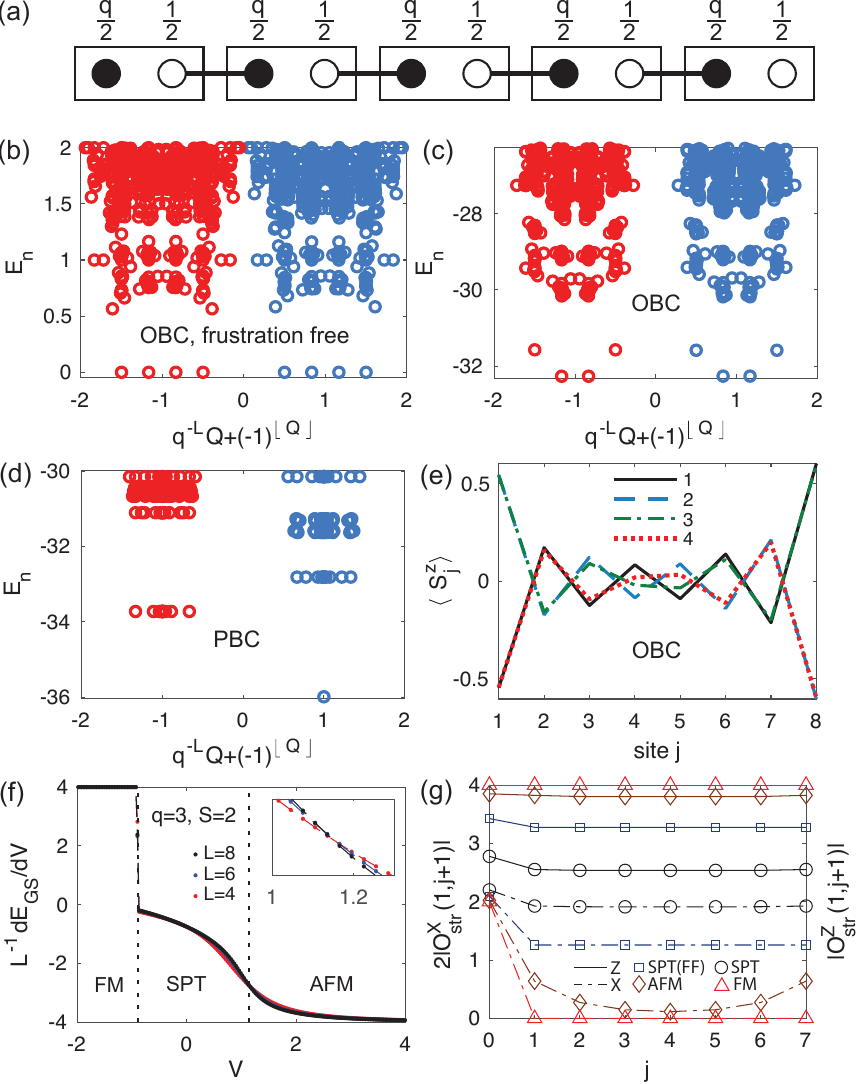}
\caption{(a) Illustration of the AKLT-like ground state. (b-g) ED results for $q=3$, $S=2$ (at $L=8$, except for (f)). (b-d) show the energy spectra of (b) $\hat{H}_\text{FF}$ in OBC, and (c-d) $\hat H$ in \cref{eq-H} at $V=0.5$, with OBC in (c) and PBC in (d). (e) $\langle \hat S_j^z\rangle$ of the $4$ OBC ground states in (c). (f) $\frac{1}{L}\frac{dE_\text{GS}}{dV}$ plotted for $\hat H$ in PBC at different $L$, and the inferred phase diagram. (g) The PBC ground state string orders (solid for $Z$, dashed for $X$), for SPT of $\hat{H}_\text{FF}$ (square) and $\hat H$ at $V=0.5$ (circle), for AFM at $V=2$ (diamond), and for FM at $V=-1.5$ (triangle).}
\label{figAKLT}
\end{figure}

\emph{The AKLT limit}---We first study a special case $S=\frac{q+1}{2}$, for which an AKLT-type symmetric ground state can be explicitly constructed. As illustrated in \cref{figAKLT}(a), by decomposing each spin $S=\frac{q+1}{2}$ into a spin $\frac{q}{2}$ and a spin $\frac{1}{2}$, we can pair the spin $\frac{1}{2}$ on site $j$ and the spin $\frac{q}{2}$ on site $j+1$ into a singlet state in the 1D irrep $R_-^{1D}$ of group $G$, which reads $\frac{1}{\sqrt{2}}\left(|\frac{1}{2},\frac{1}{2}\rangle|\frac{q}{2},-\frac{q}{2}\rangle-|\frac{1}{2},-\frac{1}{2}\rangle|\frac{q}{2},\frac{q}{2}\rangle\right)$ in their spin basis $|s,s_z\rangle$. Each site can then be projected back to spin $S=\frac{q+1}{2}$. By rewriting the spin operators with Schwinger bosons $a_j$ and $b_j$ as $S=\frac{1}{2}(a_j^\dagger a_j+b_j^\dagger b_j)$, $\hat{S}_j^z=\frac{1}{2}(a_j^\dagger a_j-b_j^\dagger b_j)$ and $\hat{S}_j^+=(\hat{S}_j^{-})^\dagger=a_j^\dagger b_j$, we can write down a unique symmetric ground state in PBC (unnormalized, sites $j\simeq j+L$):
\begin{equation}\label{eq-AKLT-GS}
|\Psi_\text{GS}\rangle=\prod_{j=1}^{L}\left[a_j^\dagger (b_{j+1}^\dagger)^q-b_j^\dagger (a_{j+1}^\dagger)^q\right]|0\rangle\ .
\end{equation}
In OBC, a spin $\frac{q}{2}$ left-edge mode and a spin $\frac{1}{2}$ right-edge mode arise, as \cref{figAKLT}(a) shows. Like the AKLT model \cite{AKLT1987}, a frustration free Hamiltonian $\hat{H}_\text{FF}$ can be constructed in this limit (see End Matter). Exact diagonalization (ED) spectra show that $\hat{H}_\text{FF}$ has a unique ground state given by \cref{eq-AKLT-GS} in PBC, and has $2(q+1)$-fold ground state degeneracy in OBC as contributed by the spin $\frac{q}{2}$ left-edge mode and the spin $\frac{1}{2}$ right-edge mode. \cref{figAKLT}(b) shows the ED spectrum of $\hat{H}_\text{FF}$ in OBC for the $q=3$, $S=2$ example. The horizontal axis is set to $q^{-L}Q+(-1)^{\lfloor Q\rfloor}$, where $Q$ is the charge sector of $\hat Q$, and $\lfloor Q\rfloor$ is its floor, to clearly differentiate two sets of ground states differing by the right-edge mode (which differ in $Q$ by $1$, colored red and blue, respectively).

The above edge modes are not fully symmetry protected away from the AKLT limit. As we shall see, the exponential symmetry group $G$ protected edge modes depend crucially on $S$ and $q$. This is rooted in the fact the Hilbert space on the $j$-th site occupies irreps of $G$ with $m_z=q^{L-j}S\ (\text{mod }1)$, which are linear if $q^{L-j}S\in\mathbb{Z}$, and projective if $q^{L-j}S\in\mathbb{Z}+\frac{1}{2}$.

\emph{SPT for integer spin $S$}---We first focus on integer spin $S\in\mathbb{Z}_+$, for which each site has a Hilbert space in a linear representation of $G$. Explicitly, we require an SPT phase to have (i) a unique gapped ground state in PBC respecting the PBC symmetry group $G_L$, and (ii) a zero-energy edge mode on each edge occupying a projective irrep of the OBC symmetry group $G$. Only such edge modes are topologically protected against any local symmetric perturbations occupying linear irreps of $G$. 

The second cohomology group $H^2(G,\text{U}(1))=\mathbb{Z}_2$ classifies the irreps of $G=\text{U}(1)\rtimes\mathbb{Z}_2$, with the group elements $0$ and $1$ being the sets of linear and projective irreps, respectively. If $G$ were a conventional symmetry group, this would imply a $\mathbb{Z}_2$ classification for 1D SPT \cite{Chen2011Classification,Chen2013GroupCohomology}, as is the case in the Haldane spin chain. However, this is not always true for the exponential symmetry group $G$. The key is to note that by \cref{eq-exp-charge}, the pseudospin $m_z$ of any irrep of $G$ on site $j$ has to be an integer multiple of $\frac{q^{L-j}}{2}$. In examining edge modes in OBC, this distinguishes the following two cases, as is summarized in \cref{table:classification}.

(i) For even $q\in 2\mathbb{Z}_+$, any irrep on sites $j<L$ must be linear. This rules out any projective irrep edge mode near the left edge $j=1$. While the right-edge site $j=L$ allows projective irreps, it cannot host a projective irrep edge mode, either, since the Hilbert space of the entire integer spin $S$ chain is in a linear representation of $G$. Therefore, no SPT exists if $q$ is even, which is consistent with the ED results (see SM \cite{suppl}).

(ii) For odd $q\in 2\mathbb{Z}_+-1$, all sites allow projective irreps, and thus the SPT classification is $H^2(G,\text{U}(1))=\mathbb{Z}_2$, like that for conventional symmetries. In this case, since group $G$ only has 2D projective irreps, each edge mode contributes a $2$-fold degeneracy, and an SPT will have $2\times 2=4$ degenerate ground states in OBC.

\begin{table}[bp]
\centering
\caption{Classification of symmetric ground states of 1D spin chains with exponential symmetry $G=\text{U}(1)\rtimes\mathbb{Z}_2$.}
\label{table:classification}
\begin{tabular}{|c|c|c|}
\hline
spin $S$ &{\ } exponential U(1) base $q${\ } & {\ }type of topology{\ } \\ 
\hline
$\mathbb{Z}_+$ & $2\mathbb{Z}_+$ & trivial  \\
\hline
$\mathbb{Z}_+$ & $2\mathbb{Z}_+-1$ & $\mathbb{Z}_2$ SPT \\
\hline
$\mathbb{Z}_+-\frac{1}{2}$ & $2\mathbb{Z}_+-1$ & LSM obstruction  \\
\hline
$\mathbb{Z}_++\frac{1}{2}$ & $2\mathbb{Z}_+$ & half-SPT  \\
\hline

\end{tabular}
\end{table}

For integer spin $S=\frac{q+1}{2}\in\mathbb{Z}_+$, for which $q$ is odd, the AKLT-type ground state in \cref{eq-AKLT-GS} is in an SPT phase. By the above reasoning, away from the AKLT limit, we expect its $(q+1)$-fold degeneracy of the spin $q/2$ left-edge mode to reduce to a $2$-fold degeneracy, while its $2$-fold degeneracy of the right-edge spin $1/2$ mode is robust. An example is the Hamiltonian $\hat{H}$ in \cref{eq-H} with $q=3$, $S=2$ at $V=0.5$, for which the ED spectra in \cref{figAKLT}(c) and (d) show $4$ degenerate ground states in OBC and $1$ ground state in PBC. \cref{figAKLT}(e) shows the site-$j$ dependence of $\langle \hat S_j^z\rangle$ of the $4$ OBC ground states, which clearly only differ at the two edges. \cref{figAKLT}(f) plots $\frac{1}{L}\frac{dE_\text{GS}}{dV}$ of the PBC ground state energy $E_\text{GS}$ of $\hat{H}$ with $q=3$, $S=2$, the $L$-dependence of which suggests a phase diagram as $V$ increases: there is a 1st order phase transition from a ferromagnetic (FM) phase to SPT at $V\approx-0.9$, and a 2nd order phase transition from SPT to an AFM phase at $V\approx 1.1$. 

\begin{figure}[t]
\centering
\includegraphics[width=3.41in]{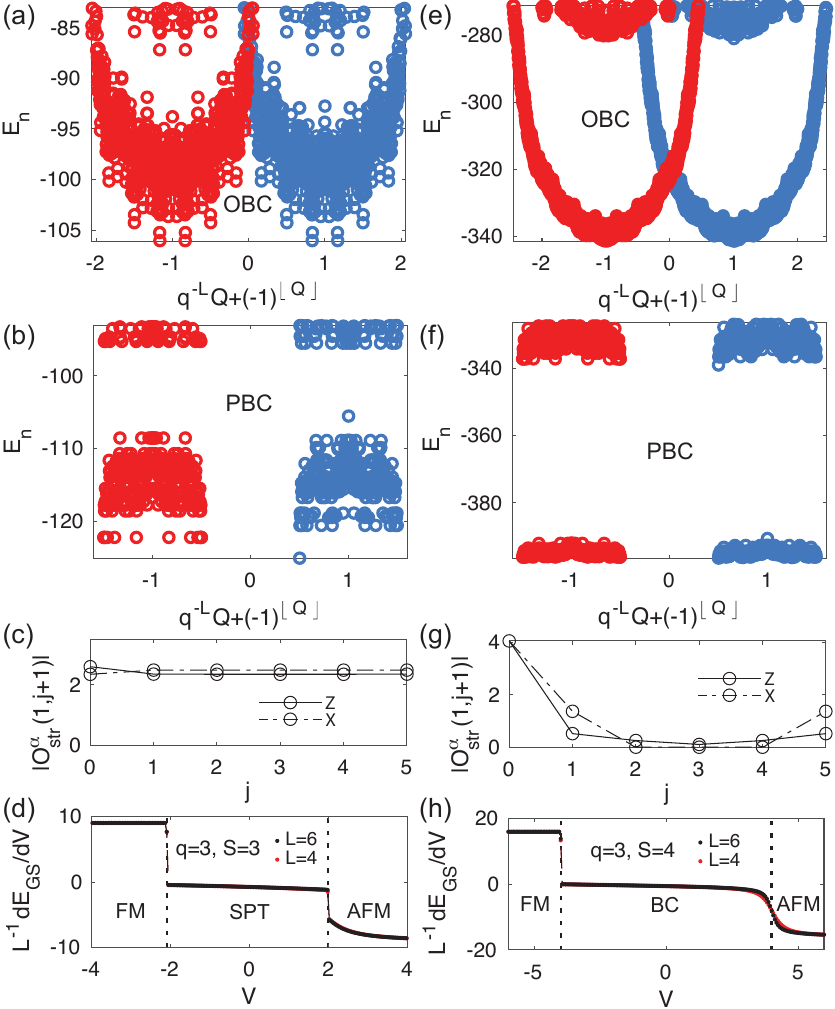}
\caption{ED results for $\hat H$ with $q=3$, $L=6$, for (a-c) $S=3$ at $V=1$, and for (e-g) $S=4$ at $V=2$. The ED spectra shown in (a),(e) for OBC, and in (b),(f) for PBC. (c) and (g) show the string orders in PBC. (d) and (h) show $\frac{1}{L}\frac{dE_\text{GS}}{dV}$ in PBC and phase diagrams for $S=3$ and $S=4$, respectively.}
\label{figSPTSSB}
\end{figure}

\begin{figure}[t]
\centering
\includegraphics[width=3.41in]{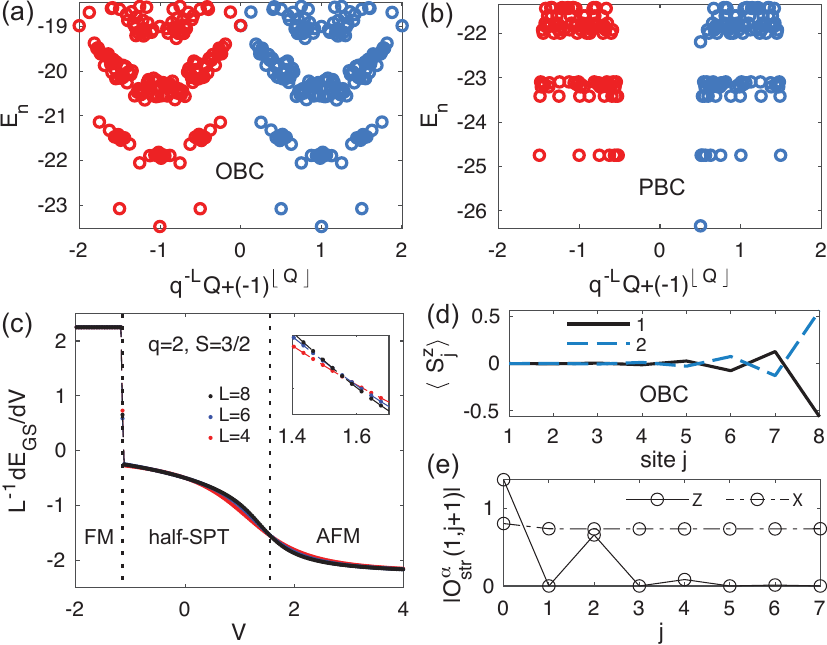}
\caption{ED results for the half-SPT phase of $\hat H$ with $q=2$, $S=3/2$, where $L=8$, $V=0.5$ except for (c). (a) and (b) show the ED spectra in OBC and PBC, respectively. (c) $\frac{1}{L}\frac{dE_\text{GS}}{dV}$ in PBC at different $L$, and the inferred phase diagram. (d) $\langle \hat S_j^z\rangle$ of the $2$ OBC ground states in (a). (e) String orders of the PBC ground state in (b). 
}
\label{figLSMSPT}
\end{figure}

\cref{figSPTSSB} shows ED results for Hamiltonian $\hat H$ for $q=3$ with higher integer spins $S$. For spin $S=3$, the system is in an SPT phase for approximately $V\in[-2,2]$, where ED shows $4$ degenerate ground states in OBC and $1$ ground state in PBC (\cref{figSPTSSB}(a),(b),(d)). For spin $S=4$ and higher, we find Hamiltonian $\hat H$ has no SPT phase (\cref{figLSMSPT}(h)). Instead, it has a breakdown condensate (BC) phase \cite{hu2025bosonic,hu2026spin} at small $V$, which is an SSB phase that breaks charge $\hat Q$ conservation. As studied in \cite{hu2026spin}, the BC phase has $q^{L}-1$ nearly degenerate ground states (one per charge sector $Q$) in PBC, and has a ground state energy $E_\text{GS}(Q)\approx E_\text{GS}(0)+Dq^{-2L}Q^2$ in charge sector $Q$ in OBC, with $D>0$. Moreover, the BC phase has no gapless bulk Goldstone mode \cite{hu2026spin,chenbt2026}, which stabilizes SSB in 1D. This exactly agrees with the ED spectra in \cref{figSPTSSB}(e)-(f), where $S=4$ and $V=2$. 

Interestingly, the $S=3$ energy levels above the SPT ground states in \cref{figLSMSPT}(a),(b) resemble the spectrum of the BC phase, implying BC is a competing phase slightly higher in energy. The BC phase is argued to resemble a disorder-free spin glass \cite{hu2026spin} due to $\mathcal{O}(q^L)$ number of ground states, while the SPT phase is viewable as a ``spin liquid" with unbroken symmetry. Thus, the $S=3$, $q=3$ SPT phase is a ``spin liquid" at exactly zero temperature, but looks like a ``spin glass" at finite temperatures approaching zero.

\emph{LSM and half-SPT for half-odd integer spin $S$}---We now examine half-odd integer spins $S\in\mathbb{Z}_+-\frac{1}{2}$. In this case, the Hilbert space contains projective irreps of $G$, and SPT cannot be defined. However, symmetric phases still show nontrivial properties. This divides into two cases below depending on $q$ (see summary in \cref{table:classification}).

(i) For odd $q\in 2\mathbb{Z}_+-1$, where every site is in a projective representation, the spin chain is subject to an LSM obstruction \cite{LSM1961,affleck1986proof}, which prohibits a unique gapped ground state preserving both translation and $G_L$ symmetries in PBC. This is consistent with the ED results (see SM \cite{suppl}). The proof is as follows. Assume $|\Psi\rangle$ is such a unique symmetric ground state in PBC. Consider a unitary symmetry operator $\hat U =\exp(i\frac{2\pi}{q^L-1}\hat Q)$, which must satisfy $\hat U|\Psi\rangle=c|\Psi\rangle$. By acting $\hat X\hat U\hat X^{-1}=\hat U^{-1}$ on $|\Psi\rangle$, one finds $c=c^{-1}$ and thus $c=\pm 1$. Also, it is easy to verify that $\hat T \hat U \hat T^{-1}=e^{i2\pi\hat S_1^z}\hat U^q=-\hat U^q$ for half-odd integer $S$, where $\hat T$ is the translation from site $j$ to site $j+1$. When acting on $|\Psi\rangle$, this gives $c=-c^q$, which contradicts with $c=\pm1$ if $q$ is odd. Thus, such a ground state $|\Psi\rangle$ cannot exist. 

(ii) For even $q\in 2\mathbb{Z}_+$, the $\hat Q$ eigenvalues are necessarily half-odd integers, so the entire chain is in a projective representation of $G$. In OBC, irrespective of symmetric or SSB, the system always has multiples of $2$ degenerate ground states, since all projective irreps of group $G$ in OBC are 2D. In PBC, however, the reduced symmetry group $G_L$ with even $q$ has two 1D irreps $\widetilde{\mathcal{R}}^\text{1D}_\pm$ reduced from the $m_z=\frac{q^L-1}{2}$ projective irrep of $G$ (see below \cref{eq-exp-charge}). In fact, group $G_L$ with even $q$ does not have projective irreps any more (see SM \cite{suppl}). This allows a symmetric gapped phase with $2$ degenerate ground states in OBC, and a unique ground state in PBC occupying one of the two 1D irreps $\widetilde{\mathcal{R}}^\text{1D}_\pm$. Since bulk properties of a phase cannot depend on boundary conditions, such a $2$-fold degeneracy in OBC must come from edge modes. In OBC, recall that for even $q$, all sites $j<L$ can only have linear irreps, and only the right-edge site $j=L$ allows projective irreps. Therefore, only the right edge can host a symmetry protected zero-energy edge mode in a projective irrep of $G$, which yields the $2$-fold degeneracy. In contrast, the left edge has no symmetry protected zero-energy edge mode. We define such a symmetric gapped phase as a ``half-SPT" phase, which has the number of edge modes halved compared to an SPT phase.

The AKLT-limit state in \cref{eq-AKLT-GS} with $S=\frac{q+1}{2}$ and $q\in2\mathbb{Z}_+$ is exactly in a half-SPT phase: only the right-edge spin $1/2$ mode is protected, while the spin $q/2$ left-edge mode is not protected. Away from the AKLT limit, \cref{figLSMSPT}(a),(b) show the ED spectra of $\hat H$ in \cref{eq-H} for $q=2$, $S=3/2$ at $V=0.5$, where we find $2$ degenerate ground states in OBC and $1$ gapped ground state in PBC. The system is in the half-SPT phase for $V\in[-1.15,1.55]$ (\cref{figLSMSPT}(c)). \cref{figLSMSPT}(d) shows that $\langle S_j^z\rangle$ of the two half-SPT ground states in OBC indeed only differ around the right edge.

\emph{String orders and entanglement spectrum}---A useful signature of 1D SPT phases is the presence of long-range string orders \cite{Nijs1989,kennedy1992,oshikawa1992hidden,PerezGarcia2008StringOrder,Pollmann2012Detection}. For $i\le j$, we define the $Z$ and $X$ string orders:
\begin{equation}\label{eq-zstring}
\begin{split}
&O^Z_\text{str}(i,j)=\langle \hat S_{i}^z \prod_{k=i}^{j-1}\exp\left(i\pi \hat S_k^z\right)\hat S_j^z\rangle \ ,\\
&O^X_\text{str}(i,j)=\langle \hat S_{i}^x \prod_{k=i+1}^{j}\exp\left(i\pi \hat S_k^x\right)\hat S_j^x\rangle\ .
\end{split}
\end{equation}
For $S\in\mathbb{Z}_+$ and $q\in 2\mathbb{Z}_+-1$, via a Kennedy-Tasaki (KT) duality similar to that of the Haldane phase \cite{kennedy1992,oshikawa1992hidden}, one can reveal a hidden $\mathbb{Z}_2\times\mathbb{Z}_2$ symmetry (see End Matter). The SPT phase in this case then maps to the SSB phase of this $\mathbb{Z}_2\times\mathbb{Z}_2$ symmetry\cite{Else2013HiddenSymmetry}, and $O^Z_\text{str}$, $O^X_\text{str}$ in \cref{eq-zstring} map to its two SSB off-diagonal long-range orders. Indeed, as shown in \cref{figAKLT}(g) and \cref{figSPTSSB}(c),(g) (calculated in PBC), the SPT phase has non-decaying $O^Z_\text{str}$ and $O^X_\text{str}$ with respect to $|j-i|$; in contrast, the FM and AFM phases have non-decaying $O^Z_\text{str}$ and decaying $O^X_\text{str}$, while the BC phase has both $O^Z_\text{str}$ and $O^X_\text{str}$ decaying.

Remarkably, although we do not find a KT-like duality for $S\in\mathbb{Z}_++\frac{1}{2}$ and $q\in 2\mathbb{Z}_+$, the half-SPT phase still has a long-range X-string order $O^X_\text{str}$ (\cref{figLSMSPT}(e)). Its $O^Z_\text{str}$ decays exponentially with $|i-j|$. Therefore, compared to the SPT phase, the half-SPT phase has half number of string orders (see understandings in End Matter).

Moreover, in both the SPT and half-SPT phases, the entanglement spectrum\cite{li2008ent,Pollmann2010EntanglementSpectrum} of the left (or equivalently the right) half of the chain is $4$-fold degenerate for the PBC ground state, and $2$-fold degenerate for the OBC ground states (see End Matter). This can be understood from the AKLT limit ground state in \cref{eq-AKLT-GS}, which can be written as a matrix product state (MPS) with bond dimension $2$ (see End Matter).

\emph{Discussion}---For spin chains respecting a continuous exponential symmetry $G=\text{U}(1)\rtimes\mathbb{Z}_2$ with integer base $q$, we showed that their symmetric gapped ground states are classified in \cref{table:classification}, depending on spin $S$ and base $q$. Particularly, we discovered a new phase named the half-SPT, which has symmetry protected zero-energy edge mode only on one end. This is distinct from any previously studied symmetric gapped phases, including those with discrete exponential symmetries \cite{Han_2023fas,Lam_classification,Kim_2025,Yao2025iia,ning2026matrixproductstatesmodulated,bulmash2025,anakru2026matrixproductstatesmodulated}. Open questions include the rigorous definition of the half-SPT phase, and the complete classification of phases with generic modulated symmetries. Experimentally, the AKLT-limit ground state as an MPS state can be prepared via quantum circuits and measurements \cite{smith2023prep,smith2024mps}. It is also promising to realize the spin Hamiltonian in \cref{eq-H} in cold atom platforms. It would be interesting to study applications of these AKLT-like states in measurement based quantum computation \cite{wei2011aklt,Else2012MeasurementBased}. 

\begin{acknowledgments}
\emph{Acknowledgments}---We thank Zhaoyu Han and Meng Cheng for helpful conversations. Z. B. thanks Amogh Anakru and Linhao Li for related collaborations. B. L. is supported by the National Science Foundation under award DMR-2141966, and National Science Foundation through the Princeton University Materials Research Science and Engineering Center DMR-2614599. Z. B. acknowledges support from National Science Foundation under award number DMR-2339319.
\end{acknowledgments}

\bibliography{exp_ref}


\section*{End Matter}

This End Matter presents the frustration free Hamiltonian and MPS representation of the AKLT limit ground state, the KT duality for understanding the string orders of SPT phase, the entanglement spectra of the SPT and half-SPT phases, and additional understandings of the half-SPT phase.

\emph{Frustration free Hamiltonian}---We derive a frustration free Hamiltonian for the AKLT-like ground state $|\Psi_\text{GS}\rangle$ in \cref{eq-AKLT-GS}, with $S=\frac{q+1}{2}$ and $q\ge 2$. The simplest exponential symmetry allowed spin flipping operator is $\hat {S}_{ j}^{+}(\hat {S}_{ j+1}^-)^q=a_j^\dagger b_j a_{j+1}^q b_{j+1}^{\dagger\ q}$ in terms of Schwinger bosons, and its Hermitian conjugate. For fixed site $j$, by rewriting \cref{eq-AKLT-GS} as $|\Psi_\text{GS}\rangle=\left(a_j^\dagger b_{j+1}^{\dagger\ q}-b_j^\dagger a_{j+1}^{\dagger\ q}\right)|\Psi_{\bar{j}}\rangle$, one can show that $\hat {S}_{ j}^{+}(\hat {S}_{ j+1}^-)^q |\Psi_\text{GS}\rangle=-b_j b_j^\dagger a_{j+1}^q a_{j+1}^{\dagger\ q} \left(a_j^\dagger b_{j+1}^{\dagger\ q} |\Psi_{\bar{j}}\rangle\right)$, and $\hat {S}_{ j}^{-}(\hat {S}_{ j+1}^+)^q |\Psi_\text{GS}\rangle=a_j a_j^\dagger b_{j+1}^q b_{j+1}^{\dagger\ q} \left(b_j^\dagger a_{j+1}^{\dagger\ q} |\Psi_{\bar{j}}\rangle\right)$. The operators $a_j^k a_j^{\dagger\ k}$, $b_j^k b_j^{\dagger\ k}$ ($k\ge 1$) can be written as functions of $\hat{S}_j^z$ when restricted to the spin $S$ Hilbert space, by which one can prove that $\hat D_{j,j+1}|\Psi_\text{GS}\rangle=0$, where
\begin{equation}
\begin{split}
&\hat D_{j,j+1}=\frac{1}{q!}\Big[(S-\hat S_j^z+1)^{-1}(qS+q\hat S_{j+1}^z+1)^{-1} \hat {S}_{ j}^{+}(\hat {S}_{ j+1}^-)^q \\
&+(S+\hat S_j^z+1)^{-1}(qS-q\hat S_{j+1}^z+1)^{-1} \hat {S}_{ j}^{-}(\hat {S}_{ j+1}^+)^q\Big]+1\ .
\end{split}
\end{equation}
One can prove that $\hat D_{j,j+1}$ commutes with the generators of group $G$. A local frustration free Hamiltonian for $|\Psi_\text{GS}\rangle$ is then
\begin{equation}\label{eq-HFF}
\hat H_\text{FF}=\sum_j \hat D_{j,j+1}^\dagger \hat D_{j,j+1}\ ,
\end{equation}
which satisfies $\hat H_\text{FF}\ge 0$ and $\hat H_\text{FF}|\Psi_\text{GS}\rangle=0$. As demonstrated in the SM \cite{suppl}, the ground state of $\hat H_\text{FF}$ with odd (even) $q$ is adiabatically connected with the SPT (half-SPT) phase of Hamiltonian $\hat H$ in \cref{eq-H}.

\emph{MPS representation}---The AKLT-limit ground state $|\Psi_{\mathrm{GS}}\rangle$ in PBC in \cref{eq-AKLT-GS}, with $S=(q+1)/2$ and $q\geq2$, can be rewritten into an MPS as
\begin{equation}
|\Psi_{\mathrm{GS}}\rangle=\sum_{\{k_j\}}\operatorname{tr}\left(\prod_{j=1}^L A^{k_j}\right)|k_1\cdots k_L\rangle ,
\end{equation}
where $|k_j\rangle$ ($-S\le k_j\le S$) is the on-site spin state satisfying $\hat S_j^z|k_j\rangle=k_j|k_j\rangle$. $A^{k}$ are $2\times 2$ (bond dimension $2$) matrices labeled by $k$, which are given by
\begin{equation}\label{eq-MPS-matrices}
\begin{split}
&A^S=\sqrt{\frac{q+1}{q+2}}\sigma^-,\qquad A^{S-1}=\sqrt{\frac{1}{q+2}}\frac{\sigma^z-\sigma^0}{2}, \\
& A^{-S}=-\sqrt{\frac{q+1}{q+2}}\sigma^+,\quad A^{-S+1}=\sqrt{\frac{1}{q+2}}\frac{\sigma^z+\sigma^0}{2},
\end{split}
\end{equation}
and $A^k=0$ for all the other $k$. Here $\sigma^0$ is the identity matrix, $\sigma^{x,y,z}$ are Pauli matrices, and $\sigma^{\pm}=\frac{1}{2}(\sigma^x\pm i\sigma^y)$. They satisfy $A^{-k}=-\sigma^xA^k\sigma^x$ and span the full algebra of $2\times2$ matrices, so the MPS is injective. The transfer matrix $\mathcal E=\sum_kA^k\otimes\overline{A^k}$ has eigenvalues $\{1,-\frac{q}{q+2},0,0\}$, which gives the correlation length $\xi=(\ln\frac{q+2}{q})^{-1}$. Consequently, the canonical parent Hamiltonian of the MPS has a unique gapped PBC ground state \cite{Fannes1992FinitelyCorrelated}. The MPS in OBC can be written using the same matrices in \cref{eq-MPS-matrices} with boundary condition vectors specified.

The local tensor satisfies the generalized $U(1)$ push-through relation
\begin{align}
e^{i2km_j\varphi}A^k=v_{j-1}(\varphi)^\dagger A^kv_j(\varphi),\quad v_j(\varphi)=e^{im_j\varphi\sigma^z},
\label{eq-pushthrough-U1}
\end{align}
where $m_j=\frac{q^{L-j}}{2}$. Thus, an internal bond of site $1\leq j\leq L-1$ carries a virtual pseudospin $m_j$, which is half-odd-integer for odd $q$ and integer for even $q$. Although \cref{eq-pushthrough-U1} holds locally for arbitrary $\varphi$, the virtual transformation closes in the PBC trace only if $(q^L-1)\varphi=2\pi n$ with $n\in\mathbb{Z}$. Indeed, with $m_0=q^L/2$ and $m_L=1/2$, this yields $v_L(\varphi)v_0(\varphi)^\dagger=e^{-i\varphi(q^L-1)\sigma^z/2}=(-1)^n\sigma^0$. This agrees with the global symmetry $G_{L}=\mathbb Z_{q^L-1}\rtimes\mathbb Z_2$ in PBC.

Lastly, since $e^{i\pi\hat S^x}|k\rangle=i^{2S}|-k\rangle$, the spin flip satisfies
\begin{equation}
\sum_{k'}\langle k'|e^{i\pi\hat S^x}|k\rangle A^{k'}=-i^{2S}\sigma^xA^k\sigma^x.
\end{equation}
Hence the virtual spin flip may be chosen as $v(X)=\sigma^x$, up to the $k$-independent phase $-i^{2S}$.

\emph{KT duality for integer $S$ and odd $q$}---For $S\in\mathbb{Z}_+$ and $q\in2\mathbb{Z}_+-1$, the Hamiltonian $\hat H$ in \cref{eq-H} has a hidden $\mathbb{Z}_2\times \mathbb{Z}_2$ symmetry similar to that of the XXZ Haldane chain. To see this, we define a KT duality unitary transformation $U_\text{KT}=\prod_{i<j}\exp\left(i\pi \hat S_{i}^z \hat S_j^x\right)$. With $S\in\mathbb{Z}_+$, all the terms $\exp\left(i\pi \hat S_{i}^z \hat S_j^x\right)$ in the product commute. This transforms the spin operators as \cite{oshikawa1992hidden} 
\begin{equation}
\begin{split}
&U_\text{KT}\hat S_j^xU_\text{KT}^{-1}=\hat S_j^x\exp\Big(i\pi \sum_{i>j}\hat S_i^x\Big)\ ,\\
&U_\text{KT}\hat S_j^zU_\text{KT}^{-1}=\exp\Big(i\pi \sum_{i<j}\hat S_i^z\Big)\hat S_j^z\ ,\\
&U_\text{KT}\hat S_j^yU_\text{KT}^{-1}=\exp\Big(i\pi \sum_{i<j}\hat S_i^z\Big)\hat S_j^y \exp\Big(i\pi \sum_{i>j}\hat S_i^x\Big)\ .\\
\end{split}
\end{equation}
Note that $\hat H$ in \cref{eq-H} only contains spin flipping terms of the form $(\hat S_j^x)^{\nu_1}(\hat S_{j+1}^x)^{\nu_2}(\hat S_j^y)^{\nu_3}(\hat S_{j+1}^y)^{\nu_4}$ with $\nu_1+\nu_2+\nu_3+\nu_4=q+1$. By definition, all matrix elements of $\hat H$ are real, thus $\nu_3+\nu_4$ is even. With $q$ odd, $\nu_1+\nu_2$ is also even. Thus, the KT transformed Hamiltonian $\hat H'=U_\text{KT} \hat H U_\text{KT}^\dagger$ remains local, and has a $\mathbb{Z}_2\times\mathbb{Z}_2$ symmetry generated by $\pi$ rotations about the $z$-axis and $x$-axis. Meanwhile, the two string orders in \cref{eq-zstring} are transformed into $O^\alpha_\text{str}(i,j)=\langle \hat S_{i}^\alpha \hat S_j^\alpha\rangle_{\hat H'}$ ($\alpha=X,Z$). Therefore, the SPT phase with long-range string orders $O^\alpha_\text{str}$ can be understood as the SSB phase of this hidden $\mathbb{Z}_2\times \mathbb{Z}_2$ symmetry, similar to the Haldane phase \cite{kennedy1992,oshikawa1992hidden}. However, the above KT transformation does not reveal a hidden symmetry for half-odd integer $S$ or even $q$.

\emph{Entanglement spectrum}---For a ground state, we derive its reduced density matrix $\rho_A$ in the left-half subregion $A=\{j|1\le j\le \lfloor L/2\rfloor\}$ of the chain, and calculate the entanglement spectrum $\zeta_n$ as the $n$-th lowest eigenvalue of the entanglement Hamiltonian $\hat H^E_A=-\ln \rho_A$. 

\begin{figure}[t!]
\centering
\includegraphics[width=3.4in]{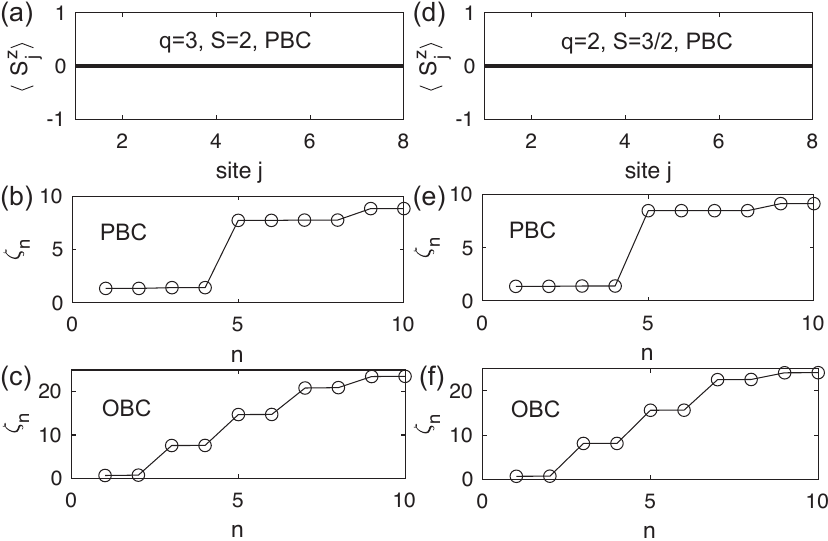}
\caption{Panels (a)-(c) (panels (d)-(f)) presents additional data for the SPT (half-SPT) phase at the parameters in \cref{figAKLT}(c)-(e) (\cref{figLSMSPT}(a)-(b)). (a) and (d) show $\langle \hat S_j^z\rangle$ of the unique PBC ground state; (b)-(f) show the ground state entanglement spectrum in PBC and OBC (labeled in each panel), respectively.
}
\label{figEmoreEE}
\end{figure}

For the SPT (half-SPT) phase of $q=3$, $S=2$ ($q=2$, $S=3/2$), \cref{figEmoreEE}(a) (\cref{figEmoreEE}(d)) shows $\langle \hat S_j^z\rangle$ of its unique PBC ground state, which is identically zero as expected. In contrast, $\langle \hat S_j^z\rangle$ for the OBC ground states in the main text \cref{figAKLT}(e) (\cref{figLSMSPT}(d)) show edge mode features. We calculate the ground state entanglement spectrum of these two phases. As shown in \cref{figEmoreEE}(b),(c),(e) and (f), the entanglement spectra of both the SPT and half-SPT phases are $4$-fold degenerate for the PBC ground state, and $2$-fold degenerate for the OBC ground states. These degeneracies are accurate up to an energy splitting $\mathcal{O}(e^{-L_A/\xi})$, where $L_A=\lfloor L/2\rfloor$ is the size of subregion $A$ and $\xi$ is the correlation length. 

The pairing in the bulk entanglement spectrum is enforced by symmetry together with the microscopic charge spectrum. Particularly, for the half-SPT phase at even $q$ and half-odd-integer $S$, every bulk interval $[i,j]$ has charge $\hat Q$ eigenvalues in $q^{L-j}(\mathbb{Z}+\tfrac12)$ and hence no neutral charge sector. In the symmetric bulk limit, its reduced density matrix commutes with both $\hat Q$ and $\hat X$, so its eigenstates can be chosen to have definite nonzero charge $Q$. The symmetry $\hat X$ maps each eigenstate to an orthogonal state of opposite charge $Q$ with the same eigenvalue, enforcing at least twofold degeneracy of every entanglement level. Thus, within the fixed microscopic Hilbert space, this pairing is symmetry protected, even though the internal virtual doublets carry linear representations of the global OBC symmetry group and the conventional projective SPT index is trivial. 

\emph{Additional notes on half-SPT}---In a finite system, the above entanglement spectrum degeneracy is only approximate up to $\mathcal{O}(e^{-L_A/\xi})$. In contrast, we note that the physical 2-fold ground state degeneracy in the half-SPT phase in OBC is exact for any system size, as is enforced by the projective representation of the entire chain. Moreover, this conclusion of half-SPT phase does not rely on describing the representation as projective. If the OBC charge $\hat Q$ changes to $q\hat{Q}$ (amounts translation by one site), all charges $Q$ become integers and non-projective (since $q$ is even), but $Q=0$ is still forbidden, and the $2$-fold degeneracy between charge sectors $Q$ and $-Q$ persists due to $\hat X \hat Q \hat X^{-1}=-\hat Q$. 

Within the original Hilbert space, symmetry enforces
at least $2$-fold degeneracy of every energy level in OBC.
In the symmetric gapped regime considered here, the
ground-state doublet is localized near the right boundary.
A symmetric finite-depth local unitary circuit can dress
this boundary degree of freedom within its finite light
cone, but cannot eliminate it. Removing this degeneracy while preserving the symmetry requires changing the microscopic Hilbert space or its symmetry representation, for example by attaching a half-odd-integer spin at the right boundary to screen the boundary doublet. Such a modification lies outside symmetry-preserving deformations of the original microscopic representation.

We also showed in \cref{figLSMSPT}(e) that the half-SPT phase has a long-range $X$-string order $O^X_\text{str}$, although we do not find a KT duality type understanding. Here we provide a direct understanding. Acting with $\prod_{k=i+1}^{j}\exp(i\pi\hat S_k^x)$ (with $i<j$) on the ground state flips the spins throughout the interval $[i+1,j]$. In the MPS picture, each local spin flip can be transferred to the neighboring virtual bonds. The virtual actions cancel pairwise in the interior, leaving only two localized changes at the endpoints of the interval. The endpoint operators in the string correlator $O^X_\text{str}$ precisely absorb these changes, so $O^X_\text{str}(i,j)$ remains finite as $|j-i|$ tends to infinity. Since the virtual charges depend on sites through powers of $q$, the required endpoint matching is itself exponentially modulated. For even $q$, these virtual charges are integer-valued, so the long-range string order $O^X_\text{str}$ can occur without a nontrivial cohomological projective SPT index, which probes the hidden entanglement structure in the bulk of the half-SPT wavefunction.

\clearpage
\onecolumngrid

\setcounter{equation}{0}
\setcounter{figure}{0}
\setcounter{table}{0}
\setcounter{page}{1}
\makeatletter
\renewcommand{\theequation}{S\arabic{equation}}
\renewcommand{\thefigure}{S\arabic{figure}}
\renewcommand{\thetable}{S\arabic{table}}
\renewcommand{\c@secnumdepth}{0}
\makeatother

\begin{center}
{\bf \large Supplemental Material}
\end{center}

This supplementary material summarizes the irreducible representations (irreps) of symmetry groups $G$ in OBC and $G_L$ in PBC, and presents additional ED numerical data.

\section{Irreducible representations of groups $G$ and $G_L$}

\subsection{Group $G$ in OBC}

The symmetry group $G=\text{U}(1)\rtimes\mathbb{Z}_2$ in OBC has group elements $g(\nu,\varphi)$ labeled by $\nu\in\{0,1\},\varphi\in(-\pi,\pi]$. The irreps of group $G$ are either 1D or 2D. There are two 1D irreps (i.e. singlet irreps) $\mathcal{R}^{1D}_\pm$, given by
\begin{equation}\label{eq-G-1D-irrep}
\mathcal{R}^\text{1D}_\pm(g(\nu,\varphi))=(\pm1)^\nu\ .
\end{equation}
Both 1D irreps are linear (i.e. non-projective) irreps, which can be viewed as pseudospin $0$ irreps. There are an infinite number of 2D irreps $\mathcal{R}^{2D}_{m_z}$ labeled by a nonzero pseudospin $m_z>0$, which reads (for $\nu\in\{0,1\},\varphi\in(-\pi,\pi]$)
\begin{equation}\label{eq-G-2D-irrep}
\mathcal{R}^\text{2D}_{m_z}(g(\nu,\varphi))= (\sigma_x)^\nu \exp\left(i m_z\varphi\sigma_z\right)\ ,
\end{equation}
where $\sigma_\mu$ ($\mu=x,y,z$) are the $2\times 2$ Pauli matrices. Since $g(1,0)g(0,\varphi)g(1,0)^{-1}=g(0,-\varphi)$, the 2D irrep must satisfy 
$\mathcal{R}^\text{2D}_{m_z}(g(1,0))\mathcal{R}^\text{2D}_{m_z}(g(0,\varphi))\mathcal{R}^\text{2D}_{m_z}(g(1,0)^{-1})=\omega(g(1,0),g(0,\varphi))\mathcal{R}^\text{2D}_{m_z}(g(0,-\varphi))$, where $\omega(g(1,0),g(0,\varphi))$ is $1$ for linear irreps, and nontrivial for projective irreps. For $\varphi=\pi$, this condition implies $e^{i2\pi m_z}=e^{-i2\pi m_z}$, and thus imposes the following constraints:
\begin{equation}
m_z\in\frac{\mathbb{Z}_+}{2}\ ,\quad \omega(g(1,0),g(0,\varphi))=\begin{cases}
1\ ,\qquad\ \  (\varphi\neq\pi)\\
(-1)^{2m_z}\ .\  (\varphi=\pi)
\end{cases}
\end{equation}
Therefore, when the pseudospin $m_z$ is an integer, the 2D irrep is linear. While when $m_z$ is half-odd integer, the 2D irrep is projective.

For irreps of $G=\text{U}(1)\rtimes\mathbb{Z}_2$, two 2D irreps with $m_z\neq m_z'$ have a tensor product decomposition $\mathcal{R}^\text{2D}_{m_z}\otimes \mathcal{R}^\text{2D}_{m_z'}=\mathcal{R}^\text{2D}_{m_z+m_z'}\oplus \mathcal{R}^\text{2D}_{|m_z-m_z'|}$; while two irreps of the same pseudospin $m_z$ have $\mathcal{R}^\text{2D}_{m_z}\otimes \mathcal{R}^\text{2D}_{m_z}=\mathcal{R}^\text{2D}_{2m_z}\oplus \mathcal{R}^\text{1D}_+\oplus \mathcal{R}^\text{1D}_-$. It is easy to check that for group $G$, two projective irreps tensor product into linear irreps, while a linear irrep and a projective irrep tensor product into projective irreps. Such a property of irreps is characterized by the second cohomology group 
\begin{equation}
H^2(G,\text{U}(1))=\mathbb{Z}_2\ ,
\end{equation}
which classifies the irreps of symmetry group $G=\text{U}(1)\rtimes\mathbb{Z}_2$ into linear (element $0$ of $\mathbb{Z}_2$) and projective (element $1$ of $\mathbb{Z}_2$).

\subsection{Group $G_L$ in PBC}

In PBC, the symmetry group is reduced to $G_L=\mathbb{Z}_{q^L-1}\rtimes\mathbb{Z}_2$, which is a subgroup of $G$. The group elements $g(\nu,\varphi)$ of $G_L$ are labeled by $\nu\in\{0,1\},\varphi\in\{\frac{2\pi n}{q^L-1}|n\in\mathbb{Z},-\frac{q^L-1}{2}<n\le \frac{q^L-1}{2}\}$. 

The discrete group $G_L$ still has the two 1D irreps $\mathcal{R}^{1D}_\pm$ given by \cref{eq-G-1D-irrep}. It also has two additional 1D irreps \begin{equation}\label{eq-G-1D-irrep-additional}
\widetilde{\mathcal{R}}^\text{1D}_\pm(g(\nu,\varphi))=(\pm1)^\nu e^{i(q^L-1)\varphi/2}\ ,
\end{equation}
which can be viewed as being reduced from the 2D irrep $\mathcal{R}^\text{2D}_{(q^L-1)/2}$ of the OBC group $G$ with pseudospin $m_z=\frac{q^L-1}{2}$. Besides, group $G_L$ has 2D irreps $\mathcal{R}^{2D}_{m_z}$ in the same form as \cref{eq-G-2D-irrep}, except that the pseudospin $m_z$ is restricted to $0<m_z< \frac{q^L-1}{2}$, $m_z\in \frac{\mathbb{Z}_+}{2}$. 

The second cohomology group of $G_L$ crucially depends on whether $q$ is odd or even:
\begin{equation}
H^2(G_L,\text{U}(1))=\begin{cases}
&\mathbb{Z}_2\ ,\qquad (q\in2\mathbb{Z}_+-1)\\
&0\ .\qquad\ \ (q\in 2\mathbb{Z}_+)
\end{cases}
\end{equation}
Accordingly, whether a 2D irrep is linear or projective is summarized in the below. Note that by definition, a 1D irrep is always linear.

(i) If $q\in2\mathbb{Z}_+-1$, a 2D irrep is linear if $m_z$ is an integer, and is projective if $m_z$ is a half-odd integer. This is the same as the case of 2D irreps of $G$, and the projective irreps are classified by $\mathbb{Z}_2$.

(ii) If $q\in2\mathbb{Z}_+$, a 2D irrep is always linear, irrespective of $m_z$.

This means when $q$ is even, a projective irrep of the OBC group $G$ will reduce to a linear irrep of the PBC group $G_L$. To see this, note that group $G_L$ has two discrete generators $\hat U =\exp(i\frac{2\pi}{q^L-1}\hat Q)$ and $\hat X$, which satisfies $\hat U^{q^L-1}=\hat X^2=1$. In a irrep $\mathcal{R}$, by adjusting the phases of their representation matrices, we can set
\begin{equation}
\mathcal{R}(\hat U)^{q^L-1}=\mathcal{R}(\hat X)^2=1\ ,
\end{equation}
where $1$ stands for the identity matrix. The only possible nontrivial phase $\omega$ that remains is
\begin{equation}
\mathcal{R}(\hat X)\mathcal{R}(\hat U)\mathcal{R}(\hat X)^{-1}=\omega \mathcal{R}(\hat U)^{-1}\ .
\end{equation}
Taking the $q^L-1$ power of both sides gives
\begin{equation}
\omega^{q^L-1}=1\ .
\end{equation}
Since $q$ is even, $q^L-1$ is odd, and thus there always exists a phase factor $\zeta$ satisfying
\begin{equation}
\zeta^{q^L-1}=1\ ,\qquad \zeta^2\omega=1\ .
\end{equation}
By redefining $\mathcal{R}'(\hat U)=\zeta\mathcal{R}(\hat U)$ and $\mathcal{R}'(\hat X)=\mathcal{R}(\hat X)$, we obtain an irrep $\mathcal{R}'$ without any phase factor in multiplicatons. Thus, for even $q$, an irrep of $G_L$ is always linear.

Particularly, for even $q$, the 2D projective irrep $\mathcal{R}^\text{2D}_{(q^L-1)/2}$ of the OBC group $G$ reduces to the two 1D linear irreps $\widetilde{\mathcal{R}}^\text{1D}_\pm(g(\nu,\varphi))$ of the PBC group $G_L$ in \cref{eq-G-1D-irrep-additional}. This is crucial for the half-SPT phase to have 2-fold degenerate ground states in OBC, and a unique groud state in PBC.

\section{Additional Numerical Calculations} 

\begin{figure}[t]
\centering
\includegraphics[width=6.8in]{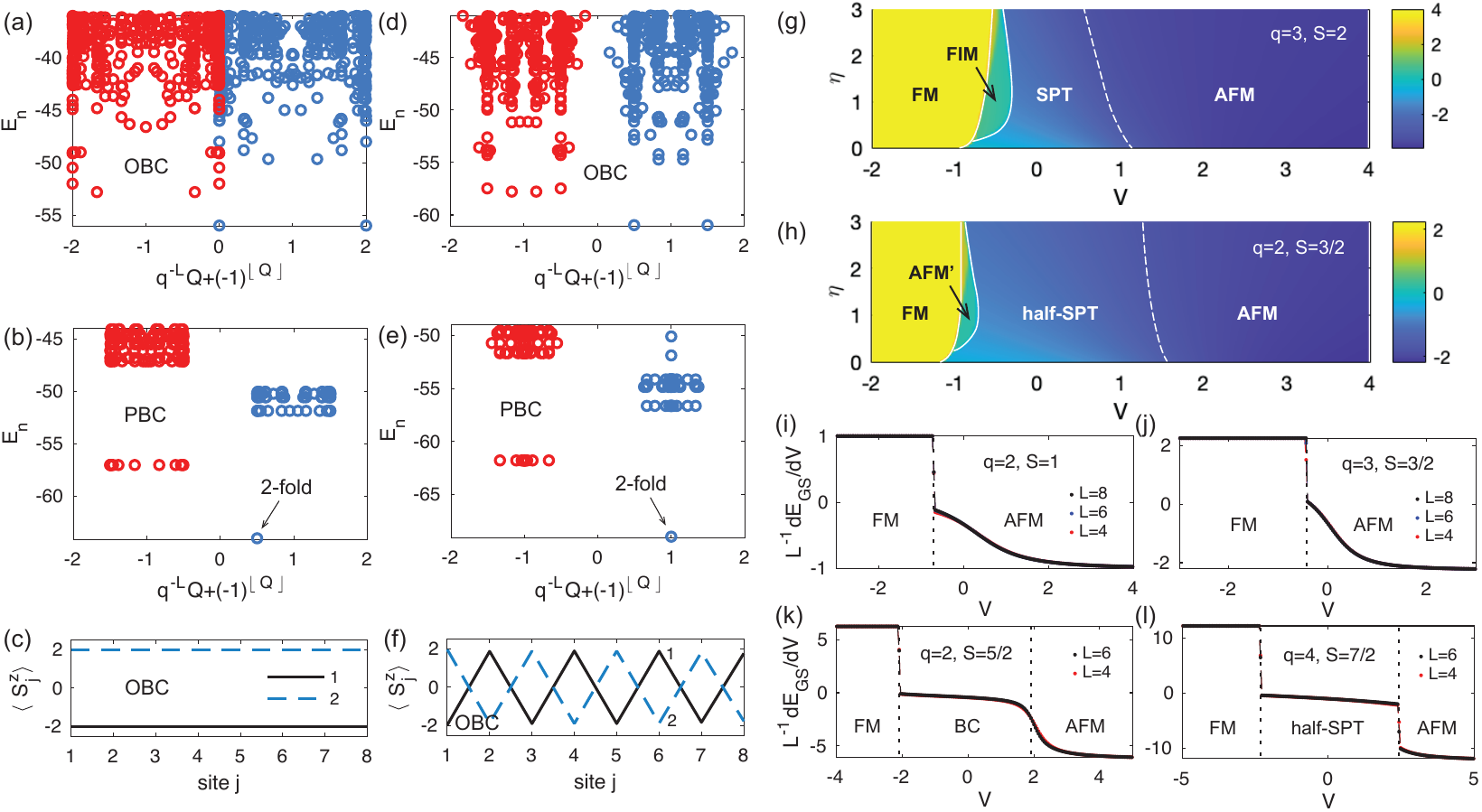}
\caption{For $\hat H$ in the main text Eq. (2) with $q=3$, $S=2$, $L=8$, panels (a)-(c) (panels (b)-(d)) show the ED energy spectrum versus charge $Q$ in OBC, in PBC, and $\langle S_j^z\rangle$ of 2 OBC ground states (labeled by $1$ and $2$), respectively, for the FM (AFM) phase at $V=-1.5$ ($V=2$). Panels (g)-(h) show the ED inferred ground state phase diagrams (via jumps or crossings of PBC ground state energy density $\frac{1}{L}\frac{dE_\text{GS}}{dV}$ at different $L$) for $\hat H_\eta$ in \cref{eq-Hinter}, with $q=3$, $S=2$ in (g), and $q=2$, $S=3/2$ in (h). FIM in (g) stands for a ferrimagnetic phase, and AFM' in (h) is an AFM phase with weaker magnetization. Solid (dashed) lines in (g)-(h) are first (second) order phase boundaries. Panels (i)-(l) shows the ED inferred phase diagrams for $\hat H$ in the main text Eq. (2) with (i) $q=2$, $S=1$, (j) $q=3$, $S=3/2$, (k) $q=2$, $S=5/2$ and (l) $q=4$, $S=7/2$. The colorbar in (g),(h) and curves in (i)-(l) give $\frac{1}{L}\frac{dE_\text{GS}}{dV}$ (for $L=8$ in (g),(h), and for different $L$ given in the legend in (i)-(l)). }
\label{figSMPD}
\end{figure}

For $\hat H$ in the main text Eq. (2), we plot the ED calculations of a typical FM state (\cref{figSMPD}(a)-(c), calculated at $V=-1.5$) and a typical AFM state (\cref{figSMPD}(d)-(f), calculated at $V=2$), where $q=3$, $S=2$ and $L=8$. Both the FM and AFM phases have 2 degenerate ground states in both OBC and PBC. $\langle \hat S_j^z\rangle$ of their $2$ OBC ground states are shown in \cref{figSMPD}(c) and (f), which clearly show the FM and AFM orders, respectively.

\cref{figSMPD}(g)-(l) shows additional phase diagrams inferred from ED calculations, with parameters given in the caption. The phase transitions are identified (subject to finite size effects) by jumps (for first order phase transitions) or crossings of different $L$ curves (for second order phase transitions) in the first order derivative of PBC ground state energy density $\frac{1}{L}\frac{dE_\text{GS}}{dV}$. 

\cref{figSMPD}(g)-(h) show the phase diagrams of Hamiltonian
\begin{equation}\label{eq-Hinter}
\hat H_\eta=\eta \hat H_\text{FF}+\sum_{j}\Big\{(1-\frac{\eta}{3})   \left[\hat {S}_{ j}^{+}(\hat {S}_{ j+1}^{-})^q + h.c.\right] +V \hat {S}_{ j}^{z}\hat {S}_{ j+1}^{z}\Big\}\ , \qquad (0\le \eta\le 3)
\end{equation}
which interpolates between $\hat H$ in the main text Eq. (2) ($\eta=0$) and the FF Hamiltonian $\hat H_\text{FF}$ in the End Matter Eq. (6) ($\eta=3$, $V=0$). The phase diagrams verified that the AKLT-limit $\hat H_\text{FF}$ ground state in the main text Eq. (3) with $q=3$, $S=2$ ($q=2$, $S=3/2$) is adiabatically connected to the SPT (half-SPT) phase of Hamiltonian $\hat H$ in the main text Eq. (2). 

\cref{figSMPD}(i)-(l) show the phase diagrams of Hamiltonian $\hat H$ in the main text Eq. (2) for other values of $S$ and $q$. \cref{figSMPD}(i) and (j) only consist of FM and AFM phases, which is consistent with the absence of nontrivial SPT in the $q\in 2\mathbb{Z}_+$, $S\in \mathbb{Z}_+$ case, and the LSM obstruction in the $q\in 2\mathbb{Z}_+-1$, $S\in \mathbb{Z}_+-\frac{1}{2}$ case, respectively. \cref{figSMPD}(k) shows that for $q=2$ and $S=5/2$ (or higher half-odd $S$), instead of half-SPT, the Hamiltonian $\hat H$ exhibits an SSB breakdown condensate (BC) phase in between the FM and AFM phases. \cref{figSMPD}(l) shows that for $q=4$, $S=7/2$, the Hamiltonian $\hat H$ exhibits a half-SPT phase between the FM and AFM phases.

\end{document}